\documentclass[twocolumn,preprintnumbers,superscriptaddress,amsmath,amssymb,aps,pra,10pt]{revtex4-1}
\usepackage{physics}
\usepackage{blindtext}
\usepackage{graphicx}
\usepackage{xcolor}
\usepackage{amsmath}
\usepackage{tabularx}
\usepackage{multirow}

\usepackage{amsthm}
\usepackage{amssymb}
\usepackage{amsfonts}
\usepackage{xr-hyper}
\usepackage{hyperref}
\usepackage{siunitx}
\hypersetup{
    colorlinks,
    linkcolor={blue},
    citecolor={blue},
    urlcolor={blue}
}
\usepackage{comment}
\usepackage{bm}
\usepackage{ulem,xpatch}
\usepackage{lineno}
\usepackage{indentfirst}
\usepackage{bbold}
\usepackage{todonotes}

\newcommand{\kb}{k_\text{B}}

\newcommand{\affil}{Photonics Laboratory, ETH Zürich, CH-8093 Zürich, Switzerland}
\newcommand{\affilQC}{Quantum Center, ETH Zurich, CH-8093 Zürich, Switzerland}

\begin{document}

\title{Trap-to-trap free falls with an optically levitated nanoparticle}

\author{M. Luisa Mattana}
\altaffiliation{Present address: Leiden Institute of Physics, Leiden University, 2300 RA Leiden, The Netherlands}
\affiliation{\affil}

\author{Nicola Carlon Zambon}
\altaffiliation{Present address: Università di Padova, Dipartimento di Fisica e Astronomia, 35134 Padova, Italy}
\affiliation{\affil}
\affiliation{\affilQC}

\author{Massimiliano Rossi}
\altaffiliation{Present address: Kavli Institute of Nanoscience, Department of Quantum Nanoscience, Delft University of Technology, 2628CJ Delft, The Netherlands}
\affiliation{\affil}
\affiliation{\affilQC}

\author{Eric Bonvin}
\affiliation{\affil}
\affiliation{\affilQC}

\author{Louisiane Devaud}
\affiliation{\affil}
\affiliation{\affilQC}

\author{Martin Frimmer}
\affiliation{\affil}
\affiliation{\affilQC}

\author{Lukas Novotny}
\affiliation{\affil}
\affiliation{\affilQC}

\begin{abstract}
We perform free-fall experiments with a charge-neutral, optically levitated nanoparticle.
This is achieved using an optical tweezer that can be rapidly toggled on and off and vertically displaced, enabling the particle to be released and recaptured after each free fall. 
The particle is insensitive to electric fields due to its charge neutrality and, during free evolution, is not subject to photon recoil heating.
We achieve free-fall durations of up to $0.25~\mathrm{ms}$ and observe a nearly two hundred-fold increase in the particle's position uncertainty at recapture.
The current limit on the free-fall time arises from the performance of the initial cooling step.
By implementing linear feedback techniques and reducing the background pressure, we expect to perform millisecond-scale free-fall experiments in ultra-high vacuum, opening new opportunities for generating large delocalizations of levitated objects.
\end{abstract}

\maketitle

\textit{Introduction} – In matter-wave experiments, objects such as atoms or molecules propagate freely through an interferometer \cite{kovachy2015quantum,Aveline2020,Fein2019}.
Free evolution increases the coherence length of the wavepacket over distances comparable to the slit separation, which is instrumental in the formation of an interference pattern \cite{Keith1991,Nairz2003}.
Typically, upon travelling through the interferometer, the object is either lost or destroyed during the detection process.
Repeating the experiment therefore relies on the ability to prepare indistinguishable copies of the object.
As we scale up to larger and more massive objects, such as nanoparticles, this indistinguishability becomes increasingly difficult to achieve \cite{Romero-Isart2011b}.
Overcoming this challenge requires repeated measurements without loss or physical destruction of the object \cite{Weiss2021,Roda-Llordes2024,Neumeier2024}.

Techniques based on levitodynamics \cite{Gonzalez-Ballestero2021}, and in particular on optical tweezers, have gained prominence for their ability to measure and control both translational and rotational degrees of freedom of sub-micrometer-sized objects, enabling ground-state preparation \cite{Delic2020, magrini2021real, ours2021, dania2024high}.
However, the ground-state de Broglie wavelength is at the picometer scale, and in order to perform matter-wave interference experiments, the state must first be delocalized.
Despite recent progress in the development of state-expansion protocols for levitated objects in purely optical \cite{Rashid2016,Duchan2024} or hybrid traps \cite{Bonvin2024, Tomassi2025}, decoherence due to background gas, photon recoil heating, and electric field noise ultimately limits the spatial coherence to subatomic distances.
This is the main reason why—so far—only two works have demonstrated a twofold increase in coherence length beyond the zero-point motion \cite{Rossi2024,Kamba2025}.

Most decoherence sources can be eliminated by allowing the particle to evolve freely in the absence of any applied potentials, under the sole influence of gravity in an Earth-based setup.
Such free-fall experiments have been demonstrated with electrically neutral particles for sensing \cite{Hebestreit2018b} and velocity measurements \cite{Kamba2023}.
However, they face a fundamental limitation: due to gravitational acceleration, the nanoparticle is displaced, eventually falls out of the trap, and is permanently lost.

Our approach to overcoming this constraint is to implement a pair of optical tweezers with tunable vertical displacement that can be rapidly toggled on and off.
The nanoparticle, initially cooled in the upper trap, is released to undergo free fall and is subsequently recaptured by activating the lower trap after a chosen evolution time.
To complete the sequence, we raise the particle back to its initial location, thereby enabling repeated cycles of free fall and recapture.

In this work, we implement such a dual optical trap.
Using a charge-neutral silica nanoparticle initialized to a cold thermal state, we perform thousands of free-fall experiments, with each lasting up to $0.25~\mathrm{ms}$.
To avoid particle loss, we optimize the recapture conditions by minimizing the particle’s energy in the lower tweezer.
During the free evolution, the particle’s position uncertainty increases.
We show that after $0.25~\mathrm{ms}$ of free fall, the state size is more than two orders of magnitude larger than the initial one.
The evolution time can be further extended by lowering the initial phonon occupation and suppressing dominant sources of decoherence.
With a ground-state-cooled nanoparticle in a UHV environment, this method is capable of generating nanometer-scale coherence lengths, paving the way for future matter-wave interference experiments in the macroscopic regime \cite{Romero-Isart2011b,Gonzalez-Ballestero2021,Neumeier2024}.

\begin{figure}[t]
		\centering
		\includegraphics[trim=0cm 0cm 0cm 0cm, width=86mm]{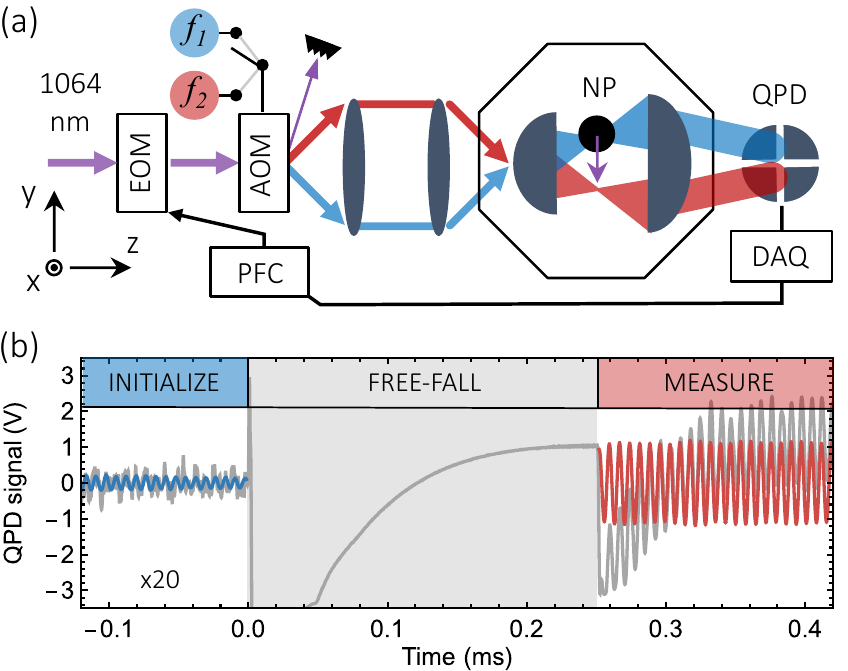}
		\caption{(a) Schematic illustration of the setup. A $1064~\mathrm{nm}$ laser enters the trapping lens at an angle controlled by the drive frequency of an acousto-optic modulator (AOM), enabling vertical displacement of the optical tweezer focus. Light forward-scattered by the nanoparticle (NP) is collected onto a quadrant photodetector (QPD) for interferometric readout of the NP position. HV, high voltage source; DAQ, data acquisition card; PFC, parametric feedback cooling electronics; EOM, electro-optical modulator. (b) Raw QPD signal (gray) from a single realization of the free fall experiment. The blue (red) line is the filtered signal before (after) a $0.25~\mathrm{ms}$ free fall showing the NP motion along the gravity axis ($y$). For clarity, signals at $t\leq 0$ have been multiplied by a factor of 20.}
		\label{fig: setup}
\end{figure}

\textit{Experiment} - A dielectric nanoparticle illuminated by a tightly focused laser experiences a restoring optical force \cite{Romero-Isart2011a}.
For small displacements, the center-of-mass (COM) of the nanoparticle undergoes harmonic evolution in three dimensions.
An essential feature of this trapping mechanism is that the optically defined spring constant (trap stiffness) can be arbitrarily tuned—or even switched off—by adjusting the laser power \cite{Millen2020}.
In this work, a silica nanoparticle with a nominal diameter of $120~\mathrm{nm}$ is optically levitated in a room-temperature vacuum chamber at a base pressure of $3\times 10^{-6}\mathrm{mBar}$.
In these conditions, the dominant decoherence mechanism is due to the background gas.
Figure~\ref{fig: setup}(a) shows the essential elements of the setup.
A laser beam (polarized along the horizontal $x$ direction, power $130~\mathrm{mW}$, wavelength $1064~\mathrm{nm}$) is focused by an aspheric lens (numerical aperture 0.75). 
The measured trap frequencies are $(\Omega_x,\Omega_y,\Omega_z)/2\pi\approx (116,141,41)~\text{kHz}$, where $y$ and $z$ denote the vertical direction (antiparallel to gravity) and the optical axis, respectively.
An acousto-optic modulator (AOM) controls both the tweezer power and the vertical deflection of the beam via the frequency $f$ of the RF driving tone.
Tilting the beam incident on the back focal plane of the lens results in a vertical displacement of the tweezer focus.
By feeding a dual-channel solid-state switch with two detuned RF frequencies, $(f_1 - f_2) = \Delta f$, we simultaneously control the tweezer status (on/off) and the tweezer position on a timescale of $300~\mathrm{ns}$.

The particle motion imprints a position-dependent phase onto the inelastically scattered laser photons.
We collect the forward-scattered light using a second collimating lens (numerical aperture 0.6) and read out the particle motion interferometrically using a quadrant photodetector (QPD).
The QPD signal is recorded by a data acquisition card (DAQ) and processed in parallel by phase-locked loops (PLLs) to track the particle's oscillation frequencies.
The PLL outputs are used to generate a parametric feedback-cooling (PFC) signal by modulating the beam intensity with an electro-optical modulator (EOM) \cite{Gieseler2012,Jain2016}.
We electrically discharge the particle using the protocol outlined in Ref.~\cite{frimmer2017controlling}.

A single realization of the free fall experiment consists of four phases.
First (Initialize): the AOM is driven at a fixed frequency $f_1$ while the nanoparticle's motion is cooled using PFC. The average COM effective temperature is reduced to $T_\text{0}^j=[12.9(2),34.1(6),42(1)]~\mathrm{mK}$ along the $j=\lbrace x,y,z\rbrace$ axes, respectively \cite{supplemental}. 
Second (Free-fall): the tweezer is switched off for a time $\tau$, during which the particle is accelerated by gravity.
Third (Measure): the tweezer is switched back on to recapture the particle after the free fall. Tuning the AOM drive frequency ($f_2$) allows overlapping the tweezer focus with the average particle position at time $\tau$. The QPD signal is recorded for $100~\mathrm{ms}$ in absence of PFC.
Fourth (Reset): the particle motion is cooled down while the AOM drive frequency is ramped back from $f_2$ to $f_1$ in $\approx1~\mathrm{s}$, thereby completing the cycle.

Figure~\ref{fig: setup}(b) shows a representative raw QPD time trace (gray line) recorded during the first three phases of the experiment for $\tau=0.25~\mathrm{ms}$. 
The signal before (after) the free fall is post-processed offline with a band-pass filter, applied forward (backward) in time.
The filter is centered on the particle's oscillation frequency and has $2~\mathrm{kHz}$ bandwidth.
See Refs.~\cite{Wiseman2009,Meng2020} for general background, and the Supplementary Materials~\cite{supplemental} for details on the filtering procedure and the filter transfer function.
The filter output provides an estimate of the particle's position $q$ and momentum $p$ along the three axes, while minimizing intrinsic measurement noise. 
We calibrate the output from voltage to displacement units following the procedure outlined in \cite{Hebestreit2018a,supplemental}.
In Fig.~\ref{fig: setup}(b) the inferred particle $y$-motion before (after) the free fall is traced with a solid blue (red) line.
In this realization of the experiment we observe a hundred-fold increase of the particle oscillation amplitude following the free evolution. 
\begin{figure}[t]
		\centering
		\includegraphics[trim=0cm 0cm 0cm 0cm, width=86mm]{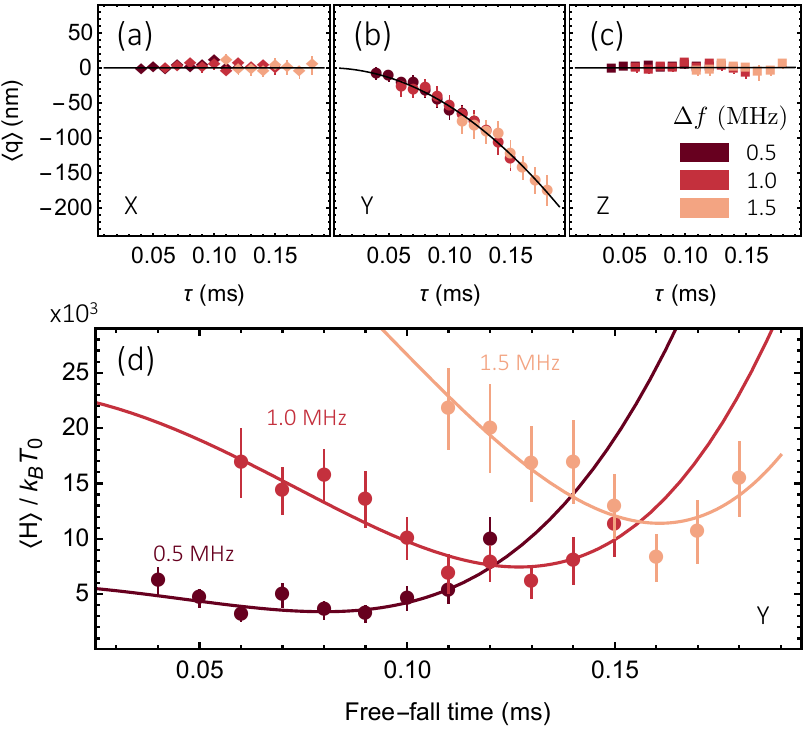}
		\caption{(a–c) Average particle center-of-mass position versus free-fall time $\tau$ along $x$ (diamonds), $y$ (circles), and $z$ (squares), respectively. Color indicates the AOM detuning value $\Delta f$ [see legend in (c)]. Solid lines are parabolic fits. (d) Energy stored in the particle motion along $y$, normalized to $k_\mathrm{B} T_0$, versus tweezer displacement. Solid lines are a joint fit to the data using Eq.~\ref{eq: recapture_energy}. In all panels, each point represents 100 experimental realizations; error bars denote $2\sigma$ confidence intervals.}
		\label{fig: ParticleEnergetics}
\end{figure}

\textit{Results} - Figure~\ref{fig: ParticleEnergetics} presents data obtained from free-fall experiments performed with varying durations $\tau$. 
Each data point is the average over one hundred realizations of the experiment.
Points in different colors correspond to different values of $\Delta f$, which determines the vertical displacement between the initialization and recapturing tweezer.
Figure~\ref{fig: ParticleEnergetics}(a-c) shows the average particle displacement $\langle q \rangle$ at recapture as a function of $\tau$ along the $x$, $y$, and $z$ axis, respectively.
The error-bars correspond to $2\sigma$ confidence intervals.
The data points representing the average displacement along $x$ and $z$ scatter around $\langle q \rangle=0$, in Figs.~\ref{fig: ParticleEnergetics}(a) and (c), while we see a clear quadratic trend along the gravitational axis $y$, in panel (b). 
Fitting a model of the uniformly accelerated motion, we estimate the acceleration $a_j$ the particle is subject to along each axis $j\in\lbrace x,y,z \rbrace$.
We obtain $a_{x}=0.0\pm 0.1~\mathrm{m/s^2}$, $a_{y}=-10.9 \pm 0.4~\mathrm{m/s^2}$ and $a_{z}=0.03\pm 0.03~\mathrm{m/s^2}$. 
Both $a_x$ and $a_z$ vanish within the error, as expected for the horizontal directions, while the small discrepancy between $a_y$ and the expected gravitational acceleration $g=9.806~\mathrm{m/s^2}$ may be ascribed to systematic errors in the volt-to-meter calibration procedure \cite{supplemental}. 
Overall, these results indicate that the particle solely evolves under the action of gravity.

Hereafter we focus on the $y$ motional degree of freedom.
We denote the normalized position and momentum of the particle along the $y$ axis by $\tilde{p}=p/p_0$ and $\tilde{q}=q/q_0$, where $q_0$ and $p_0$ are their initial root-mean-squared (rms) values under PFC.
In order to minimize the likelihood of particle loss, one needs to determine the optimal recapture conditions.
Since the depth of the optical trap is finite, reducing the loss probability is equivalent to minimizing the energy stored in the oscillator at recapture.
Figure~\ref{fig: ParticleEnergetics}(d) shows the measured average energy at recapture associated with the $y$ particle motion $\langle H_y \rangle  = \kb T_0 (\langle\tilde{p}^2\rangle+\langle\tilde{q}^2\rangle)/2$ versus the free-fall time $\tau$.
For each dataset, corresponding to three different trap-to-trap distances $d=c_\text{f}\Delta f$ ($c_\text{f}$ is a calibration factor), the energy at recapture shows a clear minimum. 
Indeed, while the kinetic energy monotonically increases with free-fall time, the potential energy can be minimized by aligning the trap center with the expectation value of the COM position at recapture, which yields the optimal trap displacement $d= g\tau^2/2$.
This explains why with increasing $d$ the minimum of $\langle H_y \rangle$ shifts towards larger free-fall times.
We jointly fit all data in Fig.~\ref{fig: ParticleEnergetics} to the model \cite{supplemental}
\begin{equation}\label{eq: recapture_energy}
	\frac{\langle H_y \rangle}{\kb T_0}=\frac{1}{2}+\frac{g^2\tau^2}{2\Omega_y^2 q_\text{0}^2} + U_0\left(1-\frac{w}{\tilde{w}}\right)\exp(-\frac{2\Delta y^2}{\tilde{w}^2}),
\end{equation}
where $\Delta y= d-g\tau^2/2=0$, $\tilde{w}^2=2 w^2+ 4 q_0^2(1+\Omega_y^2\tau^2)$, $w=0.6~\mathrm{\mu m}$ is the tweezer waist, $q_0=0.58(6)~\mathrm{nm}$ is the initial state rms motion, and $U_0$ is the potential depth in units of $\kb T_0$.
The first term in Eq.~\eqref{eq: recapture_energy} represents the particle's initial kinetic energy. 
The second term is associated with the average momentum gained during free-fall ($\langle p \rangle=-gm\tau$), and becomes comparable to the trap depth only at long times $\tau\approx 54~\mathrm{ms}$.
The third term accounts both for potential energy contributions associated with the particle-trap alignment (via $\Delta y$) and for the energy gain due to the increase in position variance resulting from free evolution (via $\tilde{w}$).
From the fit we obtain $U_0=5.4(2)\cdot 10^5$ and the
and the calibration factor $c_\text{f}=95(2)~\mathrm{nm/MHz}$ for turning an AOM detuning $\Delta f$ to a trap displacement $d$.
The fit parameters are within $10\%$ of independent estimates based on the nominal parameters of the lens system, of optical power, and of the particle properties.
Overall, we observe fair agreement of model and data, thus supporting our understanding of the particle energetics and demonstrating a systematic approach to minimizing particle loss in trap-to-trap free-fall experiments.

Next, we turn our attention to the evolution of the position and momentum variances.
For each dataset of Fig.~\ref{fig: ParticleEnergetics}, and for an additional dataset corresponding to $\tau = 0.25~\mathrm{ms}$ recorded with a different particle, we use the ensemble of trajectories to infer the state covariance matrix at recapture.
Due to the finite sampling rate of the DAQ ($0.95~\mathrm{MHz}$), the state-space ellipse associated with the inferred covariance matrix is rotated by a small angle $\delta\theta$ relative to that at the exact recapture instant.
As a result, estimates of the position and momentum marginals derived from the inferred covariance matrix may be biased.
However, since we are primarily interested in the evolution of the principal axes of the state-space ellipse (independent of $\delta\theta$) we compute the maximum (minimum) eigenvalue of the covariance matrix as $\sigma^2_q$ ($\sigma^2_p$), and define the state expansion (compression) as $\xi_q = \sigma_q / q_0$ ($\xi_p = \sigma_p / p_0$).
Introducing the mechanical damping rate due to the background gas, $\gamma = 2\pi \times 2.8(3)~\mathrm{mHz}$ at base pressure \cite{supplemental}, yields for $\gamma \tau \ll 1$
\begin{equation}\label{eq: expFac}
	\xi_q^2 \approx 1+\Omega^2_y\tau^2+\frac{2}{3}\Gamma\Omega_y^2\tau^3.
\end{equation}
Here, $\Gamma = \gamma (T_\text{env}/T_0)$, where $T_\text{env}=300~\mathrm{K}$ is the temperature of the background gas and $T_0=34~\mathrm{mK}$ is the effective temperature of the mode under PFC.
The three terms in Eq.~\eqref{eq: expFac} account for the initial rms motion $q_0$, for the growth of $\sigma_q$ at a rate proportional to $p_0$ (initial momentum is conserved in the particle’s frame), and for reheating due to the background gas, respectively.
\begin{figure}[t]
    \centering
	\includegraphics[trim=0cm 0cm 0cm 0cm, width=86mm]{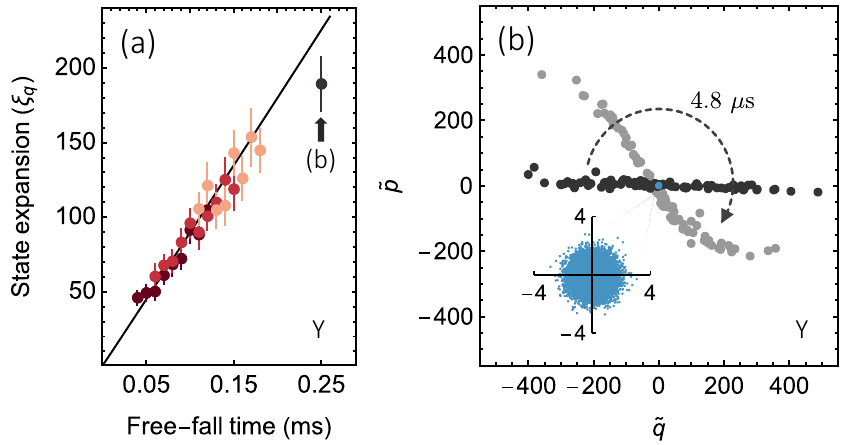}
    \caption{(a) Measured state expansion $\xi_q$ along $y$ as a function of free fall time $\tau$. Error bars correspond to $2\sigma$ standard confidence intervals. (b) Sampling of the phase-space distribution prior to the free fall (blue), at recapture (black) and after $4.8~\mathrm{\mu s}$ of evolution in the optical trap (gray) for $\tau=0.25~\mathrm{ms}$.}
    \label{fig: expansion_factor}
\end{figure}

Figure~\ref{fig: expansion_factor}(a) shows the measured state expansion $\xi_q$ as a function of the free-fall time $\tau$ for the $y$ degree of freedom. 
Additional datasets for the motion along $x$ and $z$ can be found in \cite{supplemental}.
Error bars correspond to $2\sigma$ confidence intervals.
The solid line represents the prediction from Eq.~\ref{eq: expFac}, using $\Omega_y/2\pi = 141.2~\mathrm{kHz}$—the average eigenfrequency during state initialization—and $\Gamma/2\pi = 24(3)\mathrm{Hz}$, obtained from the inferred mechanical damping rate $\gamma$ and initial temperature $T_0$.
For $\tau=0.25~\mathrm{ms}$, the measured state expansion is $\xi_q = 189(18)$, corresponding to a position standard deviation $\sigma_q=110(15)~\mathrm{nm}$.
Overall, we observe fair agreement between the data and the model, except at the longest free-fall duration.
This deviation can be attributed to sub-linear transduction in the interferometric readout of the particle displacement, which becomes significant for oscillation amplitudes comparable to a quarter of the laser wavelength.

Figure~\ref{fig: expansion_factor}(b) shows the reconstructed phase-space distributions: before the free fall (blue points), at recapture (black points), and $4.8~\mathrm{\mu s}$ later (gray points).
Position and momentum are expressed in units of the initial state uncertainties $q_0$ and $p_0$.
The phase-space distribution at recapture is strongly elongated, forming an uncertainty ellipse with semi-axes $\xi_q=189(18)$ and $\xi_p=8(4)$.
The fact that $\xi_q\xi_p>1$ is consistent with reheating due to interactions with the background gas during free-fall.
%
%
Note that after expansion, the position uncertainty $\sigma_q=110(15)~\mathrm{nm}$ becomes sufficiently large, such that Duffing nonlinearities in the optical potential can no longer be ignored.
The nonlinearity gives rise to a displacement-dependent rotation rate of the phase-space distribution, as confirmed by measurements shown in Fig.~\ref{fig: expansion_factor}(b) (gray data points).
In our current implementation, the free-fall duration is limited to $0.25~\mathrm{ms}$, beyond which we observed an increased probability of particle loss.
This is due to the fact that larger expansions result in more frequent occurrences where Duffing nonlinearities push the particle’s oscillation frequency outside the $10~\mathrm{kHz}$ bandwidth of the phase-locked loops used for PFC, see \cite{supplemental}.
As a result, the feedback loop becomes increasingly unstable and eventually fails, leading to particle loss.
Future implementations could avoid this limitation by improving the cooling performance during the initialization phase, thereby reducing the position uncertainty for a given target state expansion $\xi_q$. 
\begin{figure}[t]
		\centering
		\includegraphics[trim=0cm 0cm 0cm 0cm, width=86mm]{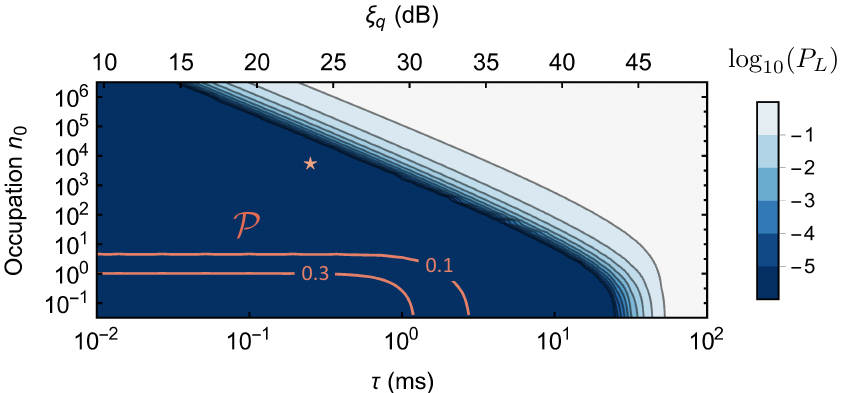}
		\caption{(a) Calculated particle loss probability ($P_L$) at the optimal recapture conditions versus  free fall time ($\tau$) and initial state occupation ($n_0$), assuming the same trap depth as in the present experiment, a base pressure of $10^{-12}~\mathrm{mbar}$ in a cryogenic environment at $5~\text{K}$. The star marks the parameter regime of the current work. Orange lines indicate the state purity $\mathcal{P}$.}
		\label{fig: FF_perspective}
\end{figure}

In the following, we discuss the capabilities of trap-to-trap free-fall protocols for generating a large quantum delocalization of the COM wavefunction, perquisite for matter–wave experiments with nanoparticles \cite{Neumeier2024}.
The key figure of merit is the coherence length $\ell=\sqrt{8}\mathcal{P}\sigma_q$, proportional to the state purity $\mathcal{P}$ and position uncertainty $\sigma_q$.
In this work, the particle was initialized in a cold, yet highly mixed thermal state.
However, ground state cooling ($n_0<1$) of a charge-neutral object can be achieved both via coherent scattering  \cite{Delic2020,Piotrowski2023} or using all-optical active feedback techniques \cite{kamba2022optical}. 
This allows preparing a high-purity initial state [with purity given by $\mathcal{P}=(2n_0+1)^{-1}$].
Preserving the state purity during free-fall, requires suppressing decoherence channels, such as those arising from interactions with background gas or black-body radiation \cite{Neumeier2024}.
This can be achieved by operating in a cryogenic ultra-high vacuum (UHV) environment \cite{Tebbenjohanns2019}.
Finally, it is necessary to reach large state expansions ($\sigma_q=\xi_q q_0$) while minimizing the particle loss probability $P_L$.

Figure~\ref{fig: FF_perspective} shows the calculated particle loss probability $P_L$ at the optimal recapture conditions as a function of the free fall time $\tau$ and initial occupation $n_0$.
We assume that the experiment is performed in a cryogenic (temperature $5~\mathrm{K}$) UHV environment reaching a base pressure $10^{-12}~\mathrm{mbar}$.
For large free fall times, particle loss occurs predominantly due to the particle position uncertainty becoming comparable with the tweezer waist $w \approx 2q_0\xi_q$. 
Due to the scaling of $q_0\propto \sqrt{n_0+1/2}$, and the expansion being linear in time for $\tau \gg \Omega^{-1}$, it is evident that low values of $n_0$ allow reaching large expansions $\xi_q$ while avoiding particle loss.
Assuming reheating during the free fall is gas dominated, we compute the state purity versus $n_0$ and $\tau$.
The result is shown in Fig.~\ref{fig: FF_perspective} using orange contour lines.
In order to retain a state purity $\mathcal{P}>0.1$ starting from $n_0=1$, the maximum free fall time is bounded to $\tau\approx 1.9~\text{ms}$.
During this time, the particle displaces due to gravity by $\langle q_y \rangle\approx 18~\mathrm{\mu m}$, which is still compatible with the geometric constraints of our trapping system.
A free evolution of this scale provides state expansion of $\xi_q\approx 1680$, and a coherence length $\ell=4.7~\mathrm{nm}$, an improvement of more than two orders of magnitude over the current state-of-the-art \cite{Rossi2024,Kamba2025}.

\textit{Conclusion} - In summary, we have demonstrated a method to recapture a nanoparticle after a free-fall experiment and recycle it for subsequent trials.
We demonstrate free fall lasting up to  $0.25~\mathrm{ms}$, corresponding to a 200-fold increase in position uncertainty ($\sigma_q=110(15)~\mathrm{nm}$), beyond which the particle loss probability increases significantly.
Longer free-fall times, thus state expansion, will become accessible by improving the cooling performance during the initialization phase, for instance, through the use of optical cold damping \cite{kamba2022optical} and better vacuum conditions.
Notably, our approach does not rely on charged particles, making it inherently robust against noise from electric fields.\\

\textbf{Acknowledgments} We thank the Q-Xtreme synergy group and our colleagues at the ETH Photonics Laboratory for fruitful discussions. M.L.M. thanks Dr. Olivier Faist for his valuable insights in the configuration of the Zurich Instruments hardware. This research has been supported by the European Research Council (ERC) under the grant agreement No. [951234] (Q-Xtreme ERC-2020-SyG), the Swiss SERI Quantum Initiative (grant no. UeM019-2) and the Swiss National Science Foundation (grant no. 51NF40-160591). N.C.Z. thanks for support through an ETH Fellowship (grant no. 222-1 FEL-30). L.D. thanks for support through a SNSF grant (no. 217122).

\bibliographystyle{apsrev4-1}
\bibliography{biblioFF.bib} 

\end{document}


\title{Supplemental Material:\\Trap-to-trap free falls with an optically levitated nanoparticle}

\author{M. Luisa Mattana}
\altaffiliation{Present address: Leiden Institute of Physics, Leiden University, 2300 RA Leiden, The Netherlands}
\affiliation{\affil}

\author{Nicola Carlon Zambon}
\altaffiliation{Present address: Università di Padova, Dipartimento di Fisica e Astronomia, 35134 Padova, Italy}
\affiliation{\affil}
\affiliation{\affilQC}

\author{Massimiliano Rossi}
\altaffiliation{Present address: Kavli Institute of Nanoscience, Department of Quantum Nanoscience, Delft University of Technology, 2628CJ Delft, The Netherlands}
\affiliation{\affil}
\affiliation{\affilQC}

\author{Eric Bonvin}
\affiliation{\affil}
\affiliation{\affilQC}

\author{Louisiane Devaud}
\affiliation{\affil}
\affiliation{\affilQC}

\author{Martin Frimmer}
\affiliation{\affil}
\affiliation{\affilQC}

\author{Lukas Novotny}
\affiliation{\affil}
\affiliation{\affilQC}

\maketitle

\section{State evolution during free-fall}\label{app: model}

In this section, we derive the main equations used to fit our data.
We start from the Langevin equations describing the system dynamics during the free-fall.
We assume that the particle is initially trapped in an optical tweezer and pre-cooled along all three motional axes.
For small displacements, the optical trap is harmonic with angular frequencies $\Omega_j$ ($j=\lbrace x,y,z\rbrace$), and the particle is initialized in a Gaussian state.
We denote with $q_\text{zpf,j}=\sqrt{\hbar/(m\Omega_j)}$ and $p_\text{zpf,j}=\sqrt{\hbar m \Omega_j}$ the rms zero-point fluctuation amplitudes of position and momentum along the $j$-th axis; $\hbar$ is the reduced Planck constant and $m$ the particle's mass.
The position and momentum variances at $t=0$ are 
\begin{equation}\label{eq: initialState_Vqp}
	\begin{aligned}
		V_{q,j}^0& = q_\text{zpf,j}^2 (n_{0,j}+1/2)\approx \frac{k_B T_0^j}{m\Omega_j^2},\\
		V_{p,j}^0&= p_\text{zpf,j}^2 (n_{0,j}+1/2)\approx m k_B T_0^j,
	\end{aligned}
\end{equation}
expressed in terms of residual occupation $n_0^{j}$ or of an effective temperature $T_0^{j}$. Notice that the right most expressions in Eq.~\ref{eq: initialState_Vqp} correspond to the classical equipartition results, which hold in the limit $n_0\gg 1$.

At $t=0$ the tweezer is switched off, and the particle state undergoes free evolution for a time $\tau$. The Langevin equations governing the evolution of the position $q_j$ and momentum $p_j$ operators read
\begin{equation}\label{eq: QLE_FF}
	\begin{aligned}
		\dot{q}_j&=p_j/m,\\
		\dot{p}_j&=-m g_j-\gamma p_j  + \xi_\text{th}(t),
	\end{aligned}
\end{equation}
where the term $g_j=(0,g,0)$ accounts for gravity, and the damping rate $\gamma$ is associated to the  fluctuating forces $\xi_\text{th}(t)$ acting on the particle due to interactions with the background gas.
The fluctuating force autocorrelation is $\langle \xi_\text{th}(t) \xi_\text{th}(t')\rangle  = 2 m k_B T \gamma \delta(t-t')$, with $T$ the temperature of the background gas.
Equation~\ref{eq: QLE_FF} can be written in a compact form introducing the state vector $\vb{v}_j=(q_j,p_j)^{T}$, a drift matrix $\vb{A}$ generating the autonomous dynamics of the system, and the drive term $\vb{w}_j(t)$ as $\dot{\vb{v}}_j=\vb{A}\vb{v}+\vb{w}_j(t)$ with
\begin{equation}
\begin{aligned}
	\vb{A}&=\begin{pmatrix}
		0 & 1/m\\
		0 & -\gamma
	\end{pmatrix},\\
	\vb{w}_j(t)&= (\xi_\text{th}(t)-m g_j)\begin{pmatrix}
		0\\
		1
	\end{pmatrix}.
\end{aligned}
\end{equation}
The matrix exponential associated with $\vb{A}$ reads
\begin{equation}
	\vb{\Phi}(t)=\exp(t\vb{A})=\begin{pmatrix}
		1 & \frac{1-e^{-\gamma t}}{m\gamma}\\
		0 & e^{-\gamma t}
	\end{pmatrix}
    .
\end{equation}
Working in the frame of the tweezer for $t\leq0$, the initial conditions are $\langle \vb{v}_j(0)\rangle=(0,0)^T$ and the expectation value of the state vector evolves according to
\begin{equation}\label{eq: ODEsol}
	\langle\vb{v}_j(t)\rangle =\vb{\Phi}(t)\langle \vb{v}_j(0)\rangle+\int_0^t \dd s  \vb{\Phi}(t-s)\langle\vb{w}_j(s)\rangle.
\end{equation} 
Since the expectation value of the fluctuating terms is zero, $\langle\vb{v}_j(t)\rangle=\langle \vb{v}_j(0)\rangle$ for $j=\lbrace x,z\rbrace$ whereas 
\begin{equation}\label{eq: stateVector_FF}
	\langle\vb{v}_y(t)\rangle=\begin{pmatrix}
		\frac{g}{\gamma^2}(1-\gamma t-e^{-\gamma t})\\
		-\frac{g m}{\gamma}(1-e^{-\gamma t})
	\end{pmatrix}\approx\begin{pmatrix}
		-\frac{g}{2}t^2\\
		-g m t
	\end{pmatrix}.
\end{equation}
In the right-hand side we have considered the limit $\gamma t\ll 1$, which is suited to describe our experiments since, at a base pressure $P=3\times 10^{-6}~\mathrm{mBar}$, and for the maximal free-fall time $\tau=0.25~\mathrm{ms}$, we have $\gamma \tau \approx 4.4\cdot 10^{-6}$. 

We now move to the nanoparticle's center of mass frame and compute the evolution of the covariance matrix elements of the system
The covariance matrix is defined as $\vb{\Sigma}_j=\langle \overline{\vb{v}_j\vb{v}_j^T}\rangle$, where the overline denotes symmetrization.
The density matrix of the initial state is $\vb{\Sigma}_{j}^0=\text{diag}(V^0_{q,j},V^0_{p,j})$, with diagonal entries given in Eq.~\ref{eq: initialState_Vqp}.
Notice that we are assuming that our mild parametric feedback does not introduce sizable correlations between position and momentum of the oscillator, which is verified a posteriori for our datasets.
In the case of cold damping, this requires the total decoherence rate to be much smaller than the oscillator eigenfrequencies $\Gamma_\text{tot}/\Omega_j\ll 1$. One finds
\begin{equation}\label{eq: CovMat_evo}
	\vb{\Sigma}_j(t)=\vb{\Phi}(t)\vb{\Sigma}_j^0\vb{\Phi}(t)^T+\int_0^t \dd s \, \vb{\Phi}(t-s)\vb{W}\vb{\Phi}(t-s)^T, 
\end{equation}
where $\vb{W}=\text{diag}(0,2mk_B T \gamma)$.
In the particle frame, the covariance matrix of all three degrees of freedom has the same form.
We therefore drop the subscript $j$.
The elements of the covariance matrix to first order in the reheating terms are
\begin{equation}\label{eq: cov_FF}
\begin{aligned}
	V_q(t)&\approx V_q^0 \left[ 1 + \Omega^2 t^2\left(1+\frac{2}{3}\frac{\Gamma t}{n_0+1/2}\right) \right],\\
	V_p(t)&\approx V_p^0 \left[1+2\frac{\Gamma t}{n_0+1/2} \right],\\
	C_{qp}(t)&\approx\sqrt{V_q^0V_p^0}\left[1+\frac{\Gamma t}{n_0+1/2}\right]\,\Omega t.
\end{aligned}
\end{equation}
where we have introduced the decoherence rate due to the background gas $\Gamma=\gamma k_B T/(\hbar\Omega)=\gamma(n_\text{th}+1/2)$.
The state purity $\mathcal{P}$ can be conveniently calculated in terms of the covariance matrix $\vb{\Sigma}_\text{r}$ associated to the reduced position $q_\text{r}=q/\sqrt{2}q_\text{zpf}$ and momentum $p_\text{r}=p/\sqrt{2}p_\text{zpf}$ as $\mathcal{P}=(4|\vb{\Sigma}_\text{r}|)^{-1/2}$; here $|\cdot|$ indicates the determinant of the covariance matrix.
Using Eq.~\ref{eq: CovMat_evo} and $\vb{\Sigma}_{0}=\text{diag}(V_0,V_0)$, with $V_0=n_0+1/2$ yields
\begin{equation}\label{eq: statePurity_FF}
	\mathcal{P}=\frac{1}{\sqrt{4V_0(V_0+2\Gamma t)+\frac{4}{3}(2V_0+\Gamma t)\Gamma\Omega^2 t^3 }},
\end{equation}
which we use to produce the results in Fig.~4 in the main text.

\section{Particle energetics}\label{app: particle energy}

After free-falling for a time $\tau$, the state of the particle is described by a Gaussian state whose expectation value and covariance matrix are given by Eq.~\ref{eq: stateVector_FF} and Eq.~\ref{eq: cov_FF}, respectively.
Clearly, during the free fall the particle will acquire momentum, and thus its kinetic energy will increase as a function of free-fall time.
Moreover, depending on the particle's position at the recapture time $\tau$ with respect to the tweezer focus, the particle may further acquire a finite potential energy in the optical trap.
Here we want to compute the expectation value of the total energy at recapture.
In vicinity of the trap focus, the conservative optical potential associated to a Gaussian beam is
\begin{equation}
\begin{aligned}
	U(\vb{q})&=U_0\left[1-\frac{\exp(-\frac{2q_x^2}{w_x^2}-\frac{2(q_y-d)^2}{w_y^2})}{1+(q_z/w_z)^2}\right]\\
	&=U_0\left[ 1-u_x(q_x)u_y(q_y)u_z(q_z)\right]
    ,
\end{aligned}
\end{equation}
where $d$ denotes the displacement of the tweezer along $y$ in the initial frame of the particle, $w_{x,y}\approx \SI{0.6}{\micro\meter}$ denote the tweezer waists along $x$ and $y$, $w_z\approx \SI{2.3}{\micro\meter}$ is the Rayleigh range of the Gaussian beam and 
\begin{equation}
	U_0=\frac{4 R_p^3 P_\text{tw}}{c w_x w_y}\frac{\epsilon_r-1}{\epsilon_r+2}
    ,
\end{equation}
with $R_p\approx 60~\mathrm{nm}$ the particle radius, $\epsilon_r\approx 2.1$ the relative dielectric permittivity of silica, $P_\text{tw}\approx 130~\text{mW}$ the tweezer power, and $c$ the speed of light in vacuum.
Since at $t=0$ the particle motion along the three axes is uncorrelated, and the free-evolution does not change this condition, the phase-space distribution of the particle can be written as $\mathcal{W}(\vb{q},\vb{p})=\prod_j\mathcal{W}_j(q_j,p_j)$ with $j=\lbrace x,y,z\rbrace$ and $\mathcal{W}_j\propto \exp[-\vb{v}_j^T\vb{\Sigma}_j\vb{v}_j]$.
The energy of the particle at recapture is
\begin{equation}
	H(\vb{q},\vb{p})=\sum_j \frac{p_j^2}{2m} + U(\vb{q})
    .
\end{equation}
Note that we neglected the gravitational potential, whose contribution over length-scales comparable to the tweezer waist is negligible if compared to the optical trap depth $U_0/(mgw_y)\approx 10^4$.
The expected value of the particle's energy at recapture is
\begin{equation}
	\begin{aligned}
		\langle H \rangle & =\int_{\mathbb{R}} \dd\vb{q}\dd\vb{p} H(\vb{q},\vb{p})\mathcal{W}(\vb{q},\vb{p})\\
		&=\langle K \rangle + U_0(1-I_x I_y I_z).
	\end{aligned}
\end{equation}
Here $\langle K \rangle = m (g\tau)^2/2 + \sum_j V_{p}^j(\tau)/(2m)$, where $V_{p}^j(\tau)$ denotes the momentum variance along the $j$-th axis at recapture, as defined in Eq.~\ref{eq: cov_FF}, and we introduced for short hand notation the quantities
\begin{equation}
	I_x =\int_{\mathbb{R}} \dd q_x\dd p_x \mathcal{W}_x u_x=\frac{w_x}{\sqrt{w_x^2+4 V_{q}^y}},
\end{equation}
\begin{equation}
	I_y =\int_{\mathbb{R}} \dd q_y\dd p_y \mathcal{W}_y u_y =\frac{w_y \exp\left( - \frac{2\Delta y^2}{w_y^2+4 V_{q}^y}\right)}{\sqrt{w_y^2+4 V_{q}^y}}
    ,
\end{equation}
\begin{equation}
 I_z=\int_{\mathbb{R}} \dd q_z\dd p_z \mathcal{W}_z u_z = \sqrt{\frac{\pi w_z^2}{2V_q^z}}\text{erfc}\left(\sqrt{\frac{w_z^2}{2V_q^z}}\right) e^{\frac{w_z^2}{2V_z}},
\end{equation}
where $\Delta y= d-g\tau^2/2$ and $V_q^j$ is given in Eq.~\ref{eq: cov_FF}.
While $I_x$ and $I_z$ are monotonous functions of the free fall time, $I_y$ also depends on the tweezer-particle separation $\Delta y$.
For a fixed free-fall time $\tau$ It is evident that the energy at recapture is minimized when $\Delta y=0$.

Since in our experiments $V_q^j/w_j^2\ll 1$, we can separate the potential energy to leading order as $\langle U\rangle\approx\langle U_x \rangle+\langle U_y \rangle+\langle U_z \rangle$ with $\langle U_j \rangle = U_0(1-I_j)$.
This approximation leads to a slight underestimation of the potential energy at recapture; however, the deviation remains below $5\%$ for the range of parameters explored in Fig.~2.
The average energy at recapture for the $y$ motion becomes
\begin{equation}
	\begin{aligned}
		\langle H_y \rangle  &  = \frac{m g^2\tau^2}{2} + \frac{V_p^y(\tau)}{2 m} + U_0[1-I_y(\Delta y)]\\
		&=k_B T_0\left[ \frac{1}{2}+ \frac{g^2\tau^2}{2V_q^0\Omega^2}+\frac{U_0}{m\Omega^2V_q^0}I_y(\Delta y) \right],
	\end{aligned}
\end{equation}
which we use to fit the data in Fig.~2(d).

\section{Particle recapture probability}\label{app: model}

Following a successful recapture, the particle evolves for $\delta t = 0.1~\mathrm{s}$ in the optical trap, while the time necessary for the particle to thermalize with the surrounding gas bath at base pressure is much longer, on the scales of $\gamma^{-1}\approx 60~\mathrm{s}$.
Therefore, the particle energy does not change significantly during the measurement step, and the recapture condition is simply that the kinetic energy of the particle at $t=\tau$ is less than the trap depth.
The kinetic energy contribution along the axes $x$ and $z$ is negligible, provided $\hbar\Omega(V_0+2\Gamma t)\ll U_0 $.
While along the gravity axis the kinetic energy increases with the square of the average momentum acquired during the free fall $\langle p_y \rangle =-m g \tau$, plus a small contribution due to  momentum spread.
The recapture condition is $p_y^2\leq 2mU(\vb{r})$.
We define the maximum momentum $p_u(\vb{r})=\sqrt{2mU(\vb{r})}$ and the $q_j$ [$p_j$] marginal of the phase space distribution as $P_{q}(q_j)$ [$P_{p}(p_j)$].
Simplifying the integrands, one finds the recapture probability
\begin{equation}\label{eq: Recap_prob}
	\begin{aligned}
		P_R&=\int \dd\vb{q} P_q (\vb{q})\int_{p_y^2\leq 2mU(\vb{r})} \dd p_y P_p(p_y)\\
		&\approx\int \dd\vb{q} \frac{P_q (\vb{q}) }{2}\left[1+\erf\left(\frac{\overline{p}_y+ p_{u}(\vb{r})}{\sqrt{2 V_p}} \right)\right]
        .
	\end{aligned}
\end{equation}
In the main text we numerically integrated Eq.~\ref{eq: Recap_prob} and defined the particle loss probability as $P_L=1-P_R$.

\section{Calibration procedure}\label{app:calibration}

\begin{figure}[t]
		\centering
		\includegraphics[trim=0cm 0cm 0cm 0cm, width=86mm]{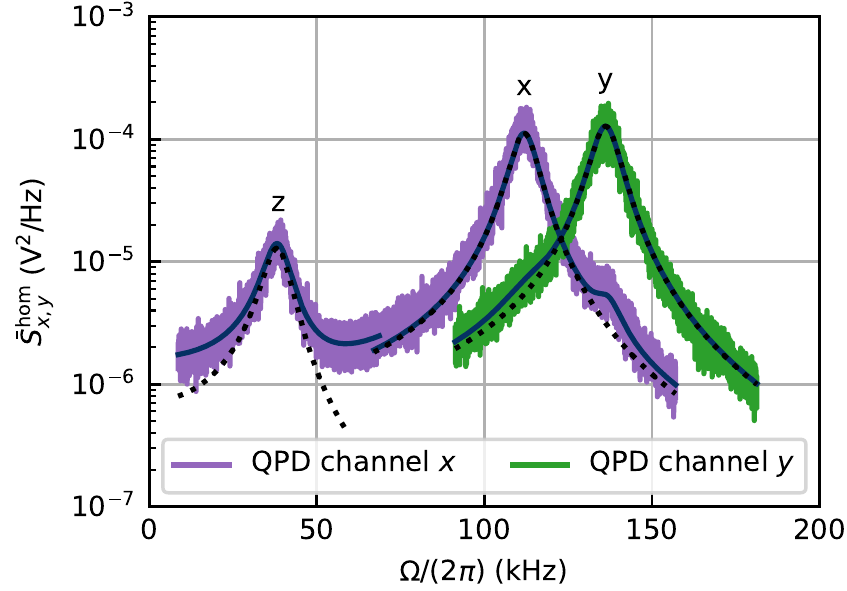}
    \caption{Power spectral density $\bar{S}^\mathrm{hom}_{x, y}$ recorded by the $x$ (horizontal, in purple) and $y$ (vertical, in green) channels of the QPD signal, acquired at \SI{9.6}{\milli\bar}. The solid blue lines show simultaneous fits of the peaks associated with the $x$, $y$ and $z$ particle motion for both QPD channels. Dotted lines represent the single oscillator model components of the global fit.
    }\label{fig:PSDhighPressures}
\end{figure}
The particle motion is encoded in the position-dependent phase imprinted onto the inelastically scattered light, which interferes with the tweezer and can be read-out interferometrically using a quadrant photodetector.
The horizontal ($x$) and vertical ($y$) motion can be efficiently detected on the corresponding differential channels of the QPD (top-down and left-right).
The longitudinal ($z$) motion modulates the global intensity of the transmitted tweezer beam, thus should not be seen on the sum channel of the QPD.
Nevertheless, imperfections in the alignment of the collection lens and in the detector electronics allows a sufficient portion of the z-motion signal to leak in both differential channels.
Here, we used the horizontal (vertical) QPD channel to read out the $x$ and $z$ ($y$) particle motion.

To convert the QPD signal from voltage to displacement units we follow the standard calibration procedure outlined in \cite{Hebestreit2018calibration}.
At high pressures, each motional degree of freedom is at thermal equilibrium with the surrounding gas at room temperature $T_\text{gas} = \SI{300}{\kelvin}$.
Using the equipartition theorem 
\begin{equation}\label{eq:equipartition}
    \frac{1}{2} k_\mathrm{B} T_\text{gas}  = \frac{1}{2} m \Omega_{ j}^2 \langle q_{j}^2 \rangle,
\end{equation}
where $\Omega_{j}$ and $\langle q_{ j}^2 \rangle$ (in units of m${^2}$) denote the eigenfrequency and displacement variance of the $j$-th translational mode, respectively.
The eigenfrequencies can be extracted from the peaks in the power spectral density (PSD) of the signal $s(t)$ recorded by the QPD.
The area under each of the peaks in the PSD  provides the variance $\langle s_j^2 \rangle$ (in voltage units) corresponding to the displacement variance for the $j$-th translational mode.
Assuming a linear relation between voltage and displacement $\langle q_j^2 \rangle= c^2_j\langle s_j^2 \rangle $ and using Eq.\eqref{eq:equipartition}, allows determining the voltage to displacement calibration factor 
\begin{equation}\label{eq:calibration}
    c_j^2 = \langle s_{j}^2\rangle \cdot \frac{m \Omega_{j}^2}{k_\mathrm{B} T_\text{gas}}
    .
\end{equation}

The linewidth $\gamma$ of the peaks in the PSD provides the mechanical damping rate of the particle.
Since the damping rate depends on the gas pressure and on particle radius $R$, it is possible to deduce $R$ from the measured damping rate \cite{Hebestreit2018calibration}
\begin{equation}
    R = 0.619 \frac{9}{\sqrt{2\pi} \rho_\mathrm{part.}} \sqrt{\frac{M}{N_A k_\mathrm{B} T_\mathrm{gas}} } \frac{P_\mathrm{gas}}{\Gamma},
\end{equation}
where $N_A$ is the Avogadro number, $\rho_\mathrm{NP} = \SI{2200}{\kilo\gram}/\SI{}{\meter\cubed}$ the density of silica, $M = \SI{28.97e-3}{\kilo\gram}/\SI{}{\mole}$ the molar mass of air, and $P_\mathrm{gas}$ the gas pressure. 
From the particle radius and the mass density $\rho_\mathrm{NP}$ one finally obtains its mass $m$.

Figure~\ref{fig:PSDhighPressures} shows the power spectral density (PSD) obtained from a 1-second time trace of the QPD signals acquired at a pressure of $\SI{9.6}{\milli\bar}$.
The particle’s motion along each axis is modelled using the transfer function of a damped harmonic oscillator, $\mathcal{L}_j(\omega) \propto |\Omega^2 - \omega^2 - i\omega\gamma|^{-2}$.
Because the tails of the peaks in the PSD partially overlap, we perform simultaneous fits of neighbouring peak pairs (solid dark lines) to more accurately estimate the relevant parameters.
From these fits, we extract the individual peak areas, linewidths $\gamma_j$, and central frequencies $\Omega_j$.
Dotted lines represent the individual oscillator components of the combined fits.
Table~\ref{tab:calibration} summarizes the results of our analysis.
\begin{table}[htb]
\centering
 \begin{tabular}{c c c c c c}
 & $\Omega_{ j}/(2\pi)$ & $c_j$ & $\gamma_j/(2\pi)$ & $m$ & $R$\\[0.5ex] 
    & [\SI{}{\kilo\hertz}] & [\SI{}{\volt}/\SI{}{\micro\meter}] & [\SI{}{\kilo\hertz}] & [\SI{}{\femto\gram}] & [\SI{}{\nano\meter}] \\ 
 \hline\hline
\multirow{3}{*}{\begin{tabular}{c|} $x$ \\ $y$ \\ $z$\end{tabular}}
 & 111.99(2) & 19(5) & 9.43(4) & 1.9(9) & 59(8) \\
 & 136.38(3) & 24(6) & 9.18(4) & 1.7(7) & 57(8) \\
 & 38.36(3) & 2.2(6)  & 8.87(4)  & 2.0(8) & 60(8) \\[1ex]
 \end{tabular}
\centering
\caption{Calibration factors $c_j$, eigenfrequencies $\Omega_{j}$ and linewidths $\gamma_j$ extracted at $P=\SI{9.6}{\milli\bar}$, for the nanoparticle used in all free fall experiments with $\tau < \SI{180}{\micro\second}$. Each degree of freedom  provides independent estimates the particle's mass $m$ and radius $R$.
}\label{tab:calibration}
\end{table}

The free fall protocols are executed at $\approx \SI{3e-6}{\milli\bar}$, after initializing each mode into a cold thermal state with effective temperature $T_\text{0, j} \ll \SI{300}{\kelvin}$ through parametric feedback cooling (PFC).
We calculate the effective temperatures by comparing the area under the PSD of each mode under PFC with the area of the peaks in the reference PSD where the particle was at thermal equilibrium with the background gas.
In the main text we also state the mechanical damping rate $\gamma$ at base pressure extrapolated from the values measured at $10~\mathrm{mBar}$, assuming a linear scaling with pressure \cite{Hebestreit2018calibration}.

\section{Data Analysis}\label{app:DataAnalysis}

The photocurrent recorded by the QPD detector contains the signal encoded in the position-dependent phase of the scattered field, along with technical noise and shot noise originating from the granular nature of the optical field.
To extract information about the particle's motion, we process the raw trajectories from each repetition of the free-fall experiment offline using a second-order bandpass filter.
This method preserves the relevant signal while suppressing excess noise outside the signal’s bandwidth.
The X (horizontal) and Y (vertical) channels of the QPD are used to infer the oscillator’s motion along the X–Z and Y axes, respectively.

In the frequency domain, the transfer function of the filter used to estimate the position signal has two poles at $\omega = 4i\Gamma_\text{f} \pm \sqrt{\Omega_\text{j}^2 - 16\Gamma_\text{f}^2}$ and one zero at $\omega = 0$.
Here, $\Omega_\text{j}$ denotes the resonance frequency of each mechanical mode, and $\Gamma_\text{f}/2\pi = 0.5~\mathrm{kHz}$. 
The filter used to extract the momentum signal shares the same poles but has no zero.
These expressions can be derived from the Kalman filter formalism in the limit where the conditional state covariance matrix $\vb{\Sigma_c} = \mathrm{diag}(V_n, V_n)$ is diagonal \cite{Wiseman2009, Meng2020}.
This limit is valid when the measurement efficiency is very low, as is the case here (approximately $0.5\%$).
We also introduce the quantity $\Gamma_\text{f} = V_n \Gamma_\text{m}$, where $\Gamma_\text{m}$ and $V_n$ denote the measurement rate and estimation noise, respectively.

To process each single-shot trajectory corresponding to one repetition of the experiment, we first compute the power spectral density (PSD) of the photocurrent before and after the free fall.
From the centroid of the peaks in the PSD, we determine the oscillation frequencies for the initialization ($\Omega_\text{init}$) and measurement ($\Omega_\text{meas}$) phases of the protocol.
These frequencies may differ due to state expansion, as the nanoparticle can explore regions where Duffing nonlinearities become significant (see Section~\ref{app:Duffing}).

Following recapture, the softening of the optical potential can induce frequency shifts as large as $5~\mathrm{kHz}$ for the y mode.
The fraction of the signal redistributed into mixing and higher-order sidebands remains on the order of a few percent.
Additionally, the oscillation amplitude—and hence the relative frequency shift—varies on timescales comparable to the inverse of the oscillator damping rate, $\gamma^{-1}$, which are much longer than the timescales relevant for inferring the oscillator’s state. 
This justifies the use of a single bandpass filter with a fixed carrier frequency for state estimation.

We then apply the filter centered at $\Omega_\text{init}$ (or $\Omega_\text{meas}$) forward (or backward) in time to the trajectory.
The forward-filtered trajectory characterizes the particle’s state before free fall, while the backward-filtered trajectory does so after free fall. 
This approach avoids artifacts arising from transients in the QPD signal during the laser on/off transitions.
We repeat this procedure independently for each axis of motion.

\begin{figure}[t]
		\centering
		\includegraphics[width=116mm]{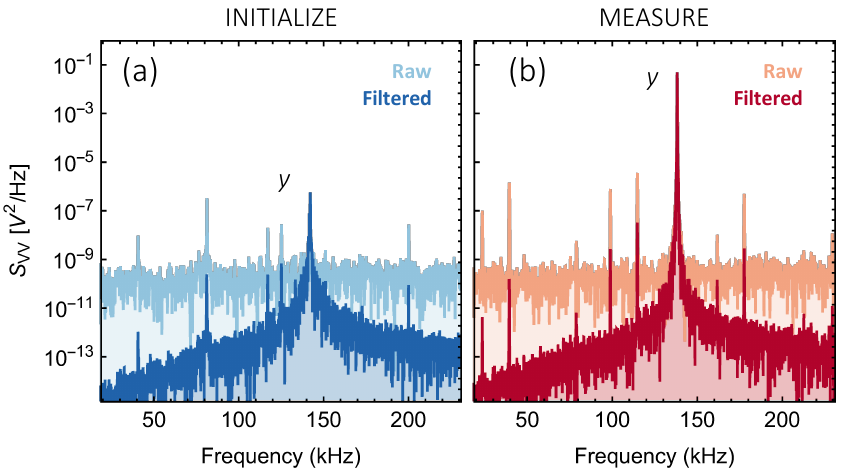}
    \caption{Power spectral density (PSD) of the raw and filtered QPD-y signal for a representative experimental realization with $\tau = 0.13~\mathrm{ms}$. Panels (a) and (b) display the PSDs recorded during the \textit{initialize} and \textit{measure} phases of the experimental sequence, respectively.}
    \label{fig:FilteredPSDs}
\end{figure}

Figure~\ref{fig:FilteredPSDs} shows the power spectrum of the raw and filtered Y signals before and after one realization of the free-fall experiment.
By ensemble-averaging the trajectories over multiple repetitions of the experiment and converting the signals from volts to meters according to the calibration procedure described in Section~\ref{app:calibration}, we obtain the mean displacement at recapture, see Fig.~2(a–c) in the main text.

Under parametric feedback cooling, the effective temperature to which the particle is initialized can fluctuate between realizations. To avoid artifacts in the analysis, we normalize all relevant quantities to those of the initial state.
From the final $1.0~\mathrm{ms}$ of the estimated position and momentum signals prior to free fall, we infer the initial state's phase-space distribution, as well as its position variance ($q_0^2$) and momentum variance ($p_0^2$). For each filtered trajectory, we then extract the particle’s position and momentum at recapture and normalize them to $q_0$ and $p_0$, respectively.

From the ensemble of realizations, we compute the second moments of the normalized position ($\tilde{q} = q/q_0$) and momentum ($\tilde{p} = p/p_0$), thereby estimating the average energy stored in the oscillator at recapture in units of the initial state energy, $k_B T_0$ (see Fig.~2(d) in the main text).
Finally, using the rescaled phase-space samples at recapture in $(\tilde{q}, \tilde{p})$ coordinates, we estimate the state covariance matrix, whose entries directly yield the parameters $\xi_q$ and $\xi_p$.
\begin{figure}[t]
		\centering
		\includegraphics[width=100mm]{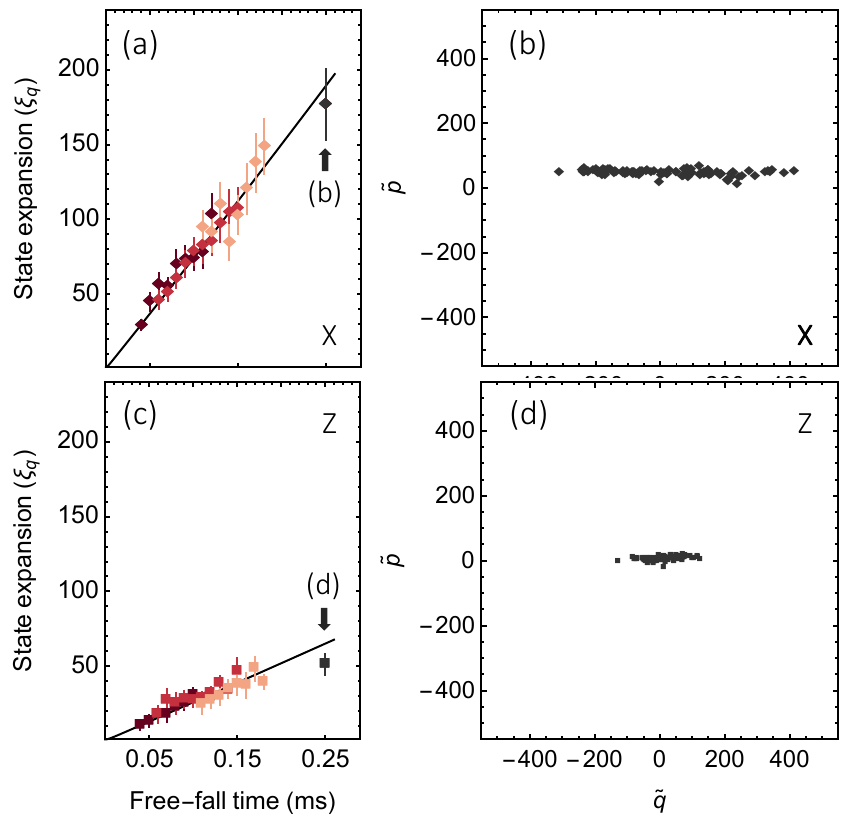}
    \caption{(a,c) Expansion factor as a function of free fall-time for the $x$ and $z$ axes respectively. The black line is the theory prediction without any free fitting parameters. (b,d) Sampling of phase-space distribution at recapture associated to the $x$ and $z$ motion of the nanoparticle, respectively.
    }
    \label{fig:Extended_datasetFig3}
\end{figure}

\section{Extended dataset}~\label{app:Extended_datasetFig3}
We present complementary results for the analysis of the expansion factor and phase space distribution of the particle along the X and Z axes of motion at recapture. 
Figure~\ref{fig:Extended_datasetFig3}(a,c) shows the measured expansion factor $\xi_q$ as a function of the free fall time for the particle motion along the $x$ and $z$ axes, respectively. 
Error bars correspond to $2\sigma$ standard confidence intervals. 
Figure~\ref{fig:Extended_datasetFig3}(b,d) shows the reconstructed phase-space distribution at recapture corresponding to the largest free fall time for the $x$ and $z$ motion, respectively.

\section{Nonlinearity of the optical trap}\label{app:Duffing}

The gradient force in an optical tweezer is approximately linear only for particle displacements that are small compared to the focal spot size.
In practice, the trapping potential is governed by the intensity profile of the laser beam, which is Gaussian.
Expanding the Gaussian potential in a Taylor series truncated at fourth order, one finds that the nonlinear restoring force acting on the trapped particle yields a displacement-dependent oscillation frequency 
\begin{equation}\label{eq:duffing}
\Omega_j = \Omega_{0, j} \left( 1 + \frac{3}{4} \sum_{i = x, y, z} \xi_{ji} \langle q_i^2 \rangle \right).
\end{equation}
For a Gaussian beam, the Duffing tensor is \cite{hebestreit2017thermal}:
\begin{equation}\label{eq:DuffingTensor}
\boldsymbol{\xi} =
\begin{pmatrix}
\xi_{xx} & \xi_{xy} & \xi_{xz} \\
\xi_{yx} & \xi_{yy} & \xi_{yz} \\
\xi_{zx} & \xi_{zy} & \xi_{zz}
\end{pmatrix}
\approx
\begin{pmatrix}
\xi_{x} & \xi_{y} & \xi_{z} \\
\xi_{x} & \xi_{y} & \xi_{z} \\
2 \xi_{x} & 2 \xi_{y} & \xi_{z}
\end{pmatrix},
\end{equation}
where $\xi_j = -2/w_j^2$ is inversely proportional to the square of the trap diameter $w_j$ along the respective axis.

After a free fall, the oscillation amplitude increases due to the gain in kinetic energy and, more significantly, due to the increased position uncertainty.
This results in a redshift of the eigenfrequencies of all three center-of-mass (COM) modes at recapture.
\begin{figure}[t]
\centering
\includegraphics[trim=0cm 0cm 0cm 0cm, width=80mm]{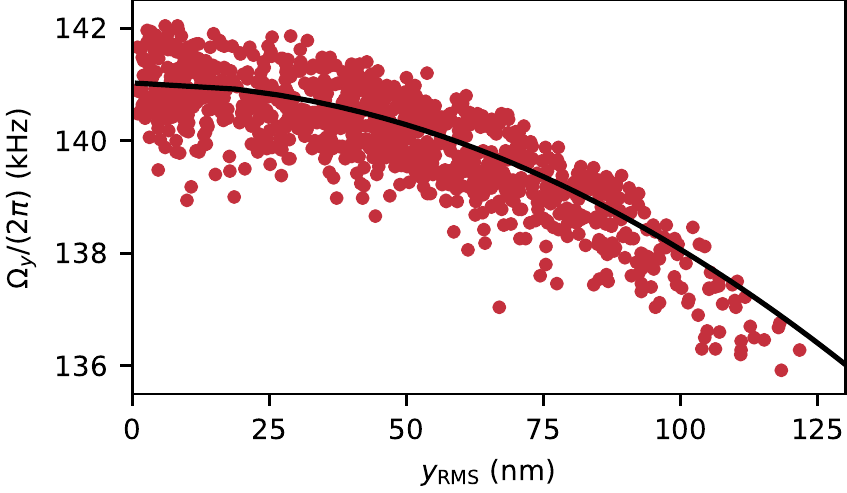}
\caption{Oscillation frequency versus root-mean-square (RMS) oscillation amplitude along the gravity axis $y$. 
The dark line is a fit to Eq.~\eqref{eq:duffing}, fixing the variances along $\langle x^2 \rangle$ and $\langle z^2 \rangle$ to their average value across all experiment repetitions.
}
\label{fig:Duffing}
\end{figure}

Figure~\ref{fig:Duffing} shows the oscillation frequency $\Omega_y$ at recapture as a function of the root-mean-square amplitude $y_\mathrm{RMS}$ along the vertical axis.
The latter is computed as the square root of the displacement variance, $y_\mathrm{RMS} = \sqrt{\langle y^2 \rangle}$.
The variance is estimated by integrating the PSD of the signal after recapture over a \SI{5}{\kilo\hertz} bandwidth around each mode, then converting the result into displacement units using the calibration factor extracted in Section~\ref{app:calibration}.
Each data point corresponds to a free-fall experiment performed between tweezer positions displaced by approximately \SI{100}{\nano\meter}, with free evolution times ranging from \SI{60}{\micro\second} to \SI{150}{\micro\second}.
As the oscillation amplitude grows to values comparable to the trap size, the eigenfrequency decreases—consistent with the softening characteristic of the Duffing nonlinearity for Gaussian beams.

The solid line in Fig.\ref{fig:Duffing} represents a fit to Eq.\eqref{eq:duffing}, treating the three tensor components $\xi_j$ as free parameters, while fixing the variances along the $x$ and $z$ axes to their average values across all experiment repetitions: $\langle q_x^2 \rangle = (\SI{38}{\nano\meter})^2$ and $\langle q_z^2 \rangle = (\SI{63}{\nano\meter})^2$.
From the fit, we extract the Duffing coefficients:
$\xi_{x} = \SI{-1.72(3)}{\micro\meter^{-2}}$,
$\xi_{y} = \SI{-2.78(2)}{\micro\meter^{-2}}$, and
$\xi_{z} = \SI{-0.32(1)}{\micro\meter^{-2}}$.
Since $\xi_j = -2/w_j^2$, we can estimate the optical beam waist and Rayleigh range:
$w_x = \SI{1.08(1)}{\micro\meter}$,
$w_y = \SI{0.85(1)}{\micro\meter}$, and
$w_z = \SI{2.50(4)}{\micro\meter}$.
The fact that $w_x > w_y$ is consistent with the linear $x$-polarization of the tweezer. All extracted values are in good agreement with those predicted by ray-tracing simulations for the nominal trapping lens geometry.

\bibliographystyle{apsrev4-1}
\bibliography{biblioFF.bib}